\documentclass[aps,twocolumn,showpacs,prl,amsmath,amssymb]{revtex4}
\usepackage{graphicx}% Include figure files
\usepackage{dcolumn}% Align table columns on decimal point
\usepackage{bm}% bold math
\usepackage{hyperref}

\newcommand{\beq}{\begin{equation}}
\newcommand{\eeq}{\end{equation}}
\newcommand{\beqa}{\begin{eqnarray}}
\newcommand{\eeqa}{\end{eqnarray}}
\newcommand{\beqar}{\begin{eqnarray*}}
\newcommand{\eeqar}{\end{eqnarray*}}
\newcommand{\bra}[1]{\mbox{$\left\langle{#1}\right|$}}
\newcommand{\ket}[1]{\mbox{$\left|{#1}\right\rangle$}}

\def\I{{\rm i}}

\def\e{{\rm e}}
\def\Tr{{\rm Tr}}

\newcounter{saveeqn}

\begin{document}

\title{Entanglement versus observables}
\author{An Min WANG}\email{anmwang@ustc.edu.cn}


\affiliation{Quantum Theory Group, Department of Modern Physics,
University of Science and Technology of China, Hefei 230026,
People's Republic of China}

\begin{abstract}
A general scheme to seek for the relations between entanglement and
observables is proposed in principle. In two-qubit systems with
enough general Hamiltonian, we find the entanglement to be the
functions of observables for six kinds of chosen state sets and
verify how these functions be invariant with time evolution.
Moreover, we demonstrate and illustrate the cases with entanglement
versus a set of commutable observables under eight kinds of given
initial states. Our conclusions show how entanglement become
observable even measurable by experiment, and they are helpful for
understanding of the nature of entanglement in physics.
\end{abstract}

\pacs{03.67.Mn, 03.65.-w}

%\keywords{}

\maketitle

\vskip -0.2in

Quantum entanglement lies at the heart of quantum mechanics and is
viewed as a useful resource in quantum information and quantum
computation \cite{Gruska}. Recently, the relations between quantum
entanglement and energy \cite{McHugh,Cavalcanti,Guhne,OurCTP} as
well as physical phenomena \cite{Brukner} have attracted a lot of
attentions from many physicists working in this area. Obviously,
these studies and their conclusions are helpful to clearly
understand the nature of quantum entanglement and effectively use it
in the practical tasks of quantum information processing.

Entanglement indeed relates closely with energy, in particular, for
spin Hamiltonian systems. However, their relation is not one to one
from our point of view. We prefer to think that this fact indicates
quantum entanglement versus quantum obsevables. This is the
motivation and propose of our study in this letter.

In fact, for a given state in a Hilbert space $\mathcal{H}$ spanned
by all $\{\ket{i}, i=1,2,\cdots\}$, its entanglement is a function
of (perhaps partial) density matrix elements $\rho_{ij}$ as follows
\beq \label{eands} E_q=f\left(\left\{\rho_{ij}\right\}\right).\eeq
When there exists a set of observables
$\{\mathcal{O}^\alpha,\alpha=1,2,\cdots\}$ in the $\mathcal{H}$, and
corresponding eigenvectors and eigenvalues of any observable
$\mathcal{O}^\alpha$ are, respectively, $\ket{v^\alpha_i}$ and
$\lambda^\alpha_i$ ($i=1,2,\cdots$). Thus, without loss of
generality, the given state can be expanded as \beq
\rho=\sum_{i,j}\rho_{ij}\ket{i}\bra{j}=\sum_{i,j}a_{ij}^\alpha
\left(\{\rho_{kl}\})\right)\ket{v^\alpha_i}\bra{v^\alpha_j}.\eeq
Hence \beq
\overline{\mathcal{O}}^\alpha=\Tr\left(\mathcal{O}^\alpha\rho\right)
=\sum_ia_{ii}^\alpha
\left(\{\rho_{kl}\})\right)\lambda^\alpha_i.\eeq If there are such a
set of observables giving an equation system made of the above form
equations with various $\alpha$ that we can solve this equation
system to obtain all $\rho_{ij}$ appearing at Eq.(\ref{eands}), that
is, \beq \rho_{ij}=g_{ij}\left(\left\{\overline{\mathcal{O}}^\alpha,
\left\{\lambda^\alpha_k\right\}\right\}\right).\eeq Thus, we can
build the relation \beq
E_q=f\left[g_{ij}\left(\left\{\overline{\mathcal{O}}^\alpha,
\left\{\lambda^\alpha_k\right\}\right\}\right)\right].\eeq This is a
general scheme to seek for the relation between entanglement and
observables in principle. It is clear that only if this relation is
true for a set or a kind of states, it will be really useful and
significant because, in the subspace of this set or this kind of
states, quantum entanglement becomes observable. Particularly, it
will be more interesting and important if this relation is invariant
with time evolution and the involved observables are commutable each
other.

Now, let we give out six sets of states of two qubits with the above
features and the following structures \beqa \rho_1&=&\left(
\begin{array}{llll}
 a_1 & 0 & 0 & \e^{-\I \alpha_1}v_1 \\
 0 & 0 & 0 & 0 \\
 0 & 0 & 0 & 0 \\
\e^{\I \alpha_1}v_1& 0 & 0 & 1-a_1
\end{array}
\right),\\
\rho_2&=&\left(
\begin{array}{llll}
 0 & 0 & 0 & 0 \\
 0 & b_2 & \e^{-\I \alpha_2}v_2 & 0 \\
 0 & \e^{\I \alpha_2}v_2 & 1-{b_2} & 0 \\
 0 & 0 & 0 & 0
\end{array}
\right),\\
\rho_3&=&\left(
\begin{array}{llll}
 0 & 0 & 0 & 0 \\
 0 & 1-{c_3}-{d_3} & \e^{-\I \alpha_3}v_3 & 0 \\
 0 & \e^{\I \alpha_3}v_3 & {c_3} & 0 \\
 0 & 0 & 0 & {d_3}
\end{array}
\right),\\
\rho_4&=&\left(
\begin{array}{llll}
 {a_4} & 0 & 0 & 0 \\
 0 & {b_4} & \e^{-\I \alpha_4}v_4 & 0 \\
 0 & \e^{\I \alpha_4}v_4 & 1-{a_4}-{b_4} & 0 \\
 0 & 0 & 0 & 0
\end{array}
\right),\\
\rho_5&=&\left(
\begin{array}{llll}
 1-{c_5}-{d_5} & 0 & 0 & \e^{-\I \alpha_5}v_5 \\
 0 & 0 & 0 & 0 \\
 0 & 0 & {c_5} & 0 \\
 \e^{\I \alpha_5}v_5 & 0 & 0 & {d_5}
\end{array}
\right),\\
\rho_6&=&\left(
\begin{array}{llll}
 1-{b_6}-{d_6} & 0 & 0 & \e^{-\I \alpha_6}v_6 \\
 0 & {b_6} & 0 & 0 \\
 0 & 0 & 0 & 0 \\
 \e^{\I \alpha_6}v_6 & 0 & 0 & {d_6}
\end{array}
\right),\eeqa where we have used the Hermi and unit trace properties
of the density matrix. As to the positive conditions of them are
also required. In order to clearly determine their entanglement by
negativity measure \cite{n1,n2}, we assume all diagonal elements and
$v_i$ in these density matrices are not negative (negative $v_i$ can
be rewritten by absorbing its minus sign into $\e^{\pm\I\alpha_i}$).
It is easy to calculate their negativity \beq\begin{array}{ll}
N_1=\left|v_1\right|,&\quad
N_2=\left|v_2\right|,\\[6pt] N_3=\sqrt{d_3^2+v_3^2}-d_3,&\quad
N_4=\sqrt{a_4^2+v_4^2}-a_4,\\[6pt] N_5=\sqrt{c_5^2+v_3^2}-c_5,&\quad
N_6=\sqrt{b_6^2+v_4^2}-b_6.
\end{array}\eeq

Introduce the spin tensor for two qubit systems \beq
s_{\mu\nu}=\sigma_\mu\otimes\sigma_\nu \quad (\mu,\nu=0,1,2,3),\eeq
where $\sigma_0$ is $2\times 2$ identity matrix and $\sigma_i$
$(i=1,2,3)$ are the Pauli matrices. Thus, the total spin vector and
its square read \vskip -0.1in \beqa
S_i&=&\frac{1}{2}\left(s_{i0}+s_{0i}\right),\quad S^2=\sum_{i=1}^3
S_i S_i.\eeqa Here, $\hbar$ is taken as 1 for simplicity. Denoting
the expected value of an operator $\mathcal{O}$ in the state
$\rho_i$ by $\langle{\mathcal{O}}\rangle_i$, we can express the
entanglement by the some components of spin tensor, that is \vskip
-0.2in \beqa\label{nsr1}
N_1&=&\frac{1}{2}\sqrt{(\langle{s}_{11}\rangle_1^2+(\langle{s}_{12}\rangle_1^2},\\
\label{nsr2}
N_2&=&\frac{1}{2}\sqrt{\langle{s}_{11}\rangle_2^2+\langle{s}_{12}\rangle_2^2},\\
\label{nsr3}
N_3&=&\frac{1}{2}\sqrt{\langle{s}_{11}\rangle_3^2+\langle{s}_{12}\rangle_3^2
+\langle{S_z}\rangle_3^2}+\langle{S_z}\rangle_3,\\
\label{nsr4}
N_4&=&\frac{1}{2}\sqrt{\langle{s}_{11}\rangle_4^2+\langle{s}_{12}\rangle_4^2
+\langle{S_z}\rangle_4^2}-\langle{S_z}\rangle_4,\\
\label{nsr5}
N_5\!&=&\!\frac{1}{2}\sqrt{\langle{s}_{11}\rangle_5^2\!+\!\langle{s}_{12}\rangle_5^2
\!+\!\left(2\!-\!\langle{S^2}\rangle_5\right)^2}\!+\!\langle{S^2}\rangle_5\!-\!
2, \hskip 1cm\\
\label{nsr6}
N_6\!&=&\!\frac{1}{2}\sqrt{\langle{s}_{11}\rangle_6^2\!+\!\langle{s}_{12}\rangle_5^2
\!+\!\left(2\!-\!\langle{S^2}\rangle_6\right)^2}\!+\!\langle{S^2}\rangle_6\!-\!
2.\eeqa Therefore, the entanglement, for the given six kinds of
state sets, can be expressed as the functions of some observables.

It is more interesting how these relations evolute with time.
Without loss of generality, the general Hamiltonian of two-qubit
systems can be written as \beq
H=\sum_{\mu,\nu=0}^3h_{\mu\nu}\sigma_\mu\otimes\sigma_\nu=\sum_{\mu,\nu=0}^3h_{\mu\nu}s_{\mu\nu}.\eeq
where $h_{\mu\nu}$ are real. Based on the facts that Hamiltonian is
made of the spin tensor and that entanglement can expressed as the
spin tensor, it is not surprised that there exists the relation
between entanglement and energy. However, entanglement quantity and
its evolution with time can not be well determined and described by
only using energy from our point of view.

It is easy to verify the sufficient conditions to keep the form
invariance of time evolution of the above relations
(\ref{nsr1},\ref{nsr2}) under the following two kinds form of
Hamiltonians \beqa\label{h1}
H[1]&=&h_{30}s_{30}+h_{03}s_{03}+f_1\left(s_{12}-s_{21}\right)\nonumber\\
& &+g_1\left(s_{11}+s_{22}\right)+h_{33}s_{33},\\
\label{h2} H[2]&=&h_{30}s_{30}+h_{03}s_{03}+f_2\left(s_{12}+s_{21}\right)\nonumber\\
& &+g_2\left(s_{11}-s_{22}\right)+h_{33}s_{33}. \eeqa In fact, the
verification process is standard and simple. We first expand the
density matrix by using the basis made of the eigenvectors of chosen
Hamiltonian and their conjugate vectors. Then, acting the time
evolution operator (independent of time) on it, we can obtain the
final density matrix at a given time. It is key matter to make the
final state has the same structure as some $\rho_i$. Finally, we
calculate its negativity. It is clear that we chose such
Hamiltonians (\ref{h1}, \ref{h2}) and following their
simplifications (\ref{h21})--(\ref{h12}) in order to guarantee the
required structure of final state. This is arrived at because these
chosen Hamiltonians have appropriate eigenvectors and eigenvalues.
The details is omitted in order to save space.

Similarly, we can verify that the relation (\ref{nsr3},\ref{nsr4})
and (\ref{nsr5},\ref{nsr6}) have the form invariance of time
evolution, respectively, under $H[1]$ and $H[2]$.

It is more significant if, in the above the relations between
entanglement and observables, the involved observables are
commutable each other. Actually, this can be obtained by taking
particular initial states. Firstly, let us set the initial state as
\beq \ket{\psi}=\sin\theta_1\ket{00}+\e^{-\I
\alpha_1}\cos\theta_1\ket{11}.\eeq Obviously, it is entangled and
will evolute to the first kind $\rho_1$. Its negativity reads \beq
N_\psi(t)=\frac{1}{2}\sqrt{1-\langle{S}_z\rangle_2(t)}.\eeq It is
both formally invariant and numerically conversed for the time
evolution with $H[1]$, that is, $N_\psi(0)=N_\psi(t)$. While under
$H[2]$, it only keeps invariance in form, but the conversation is
broken in general. In order to understand how entanglement evolute,
we chose two simplified forms of $H[2]$ \beqa\label{h21}
H[2,1]&=&\frac{1}{2}\omega_2\left(\sigma_3\otimes\sigma_0-\sigma_0\otimes\sigma_3\right)
\nonumber\\
& &+g_2\left(\sigma_1\otimes\sigma_1-\sigma_2\otimes\sigma_2\right)+h_2\sigma_3\otimes\sigma_3,\\
\label{h22}
H[2,2]&=&\frac{1}{2}\omega_2\left(\sigma_3\otimes\sigma_0-\sigma_0\otimes\sigma_3\right)
\nonumber\\
&
&+f_2\left(\sigma_1\otimes\sigma_2+\sigma_2\otimes\sigma_1\right)+h_2\sigma_3\otimes\sigma_3.
\eeqa Then, let us fix $\alpha_1$, for example, $\alpha_1=0$, we can
draw Fig.1 and Fig.2 to show the surface graphics determined by
$\theta$, $\langle S_z\rangle(t)$ and $N(t)$.
\begin{figure}[h]%[tbp]
\begin{center}
\includegraphics[scale=0.56]{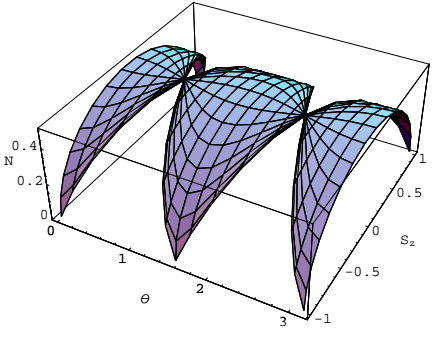}
\end{center}
\vskip -0.1in \caption{Under $H[2,1]$, $\langle
S_z\rangle_\psi(t)=-\cos\left(2T\right)\cos\left(2\theta_1\right)$,
$N_\psi(t)=\sqrt{1-\cos^2\left(2T\right)\cos^2\left(2\theta_1\right)}/2$
($\alpha_1=0$, $T=2g_2$).} \label{mypic1}
\end{figure}
\begin{figure}[h]%[tbp]
\begin{center}
\includegraphics[scale=0.56]{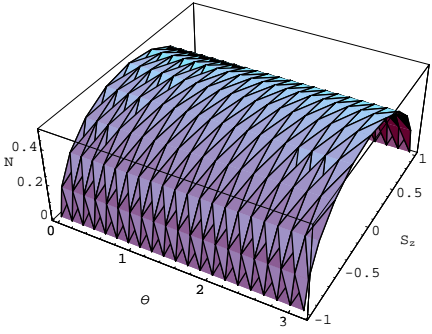}
\end{center} \vskip -0.2in
\caption{Under $H[2,2]$, $\langle
S_z\rangle_\psi(t)=-\cos\left(2T-2\theta_1\right)$,
$N_\psi(t)=\sqrt{\sin^2\left(2T-2\theta_1\right)}/2$ ($\alpha_1=0$,
$T=2f_2$).}\label{mypic2}
\end{figure}

When the initial state is taken as \beq
\ket{\phi}=\sin\theta_2\ket{01}\pm\cos\theta_2\ket{10},\eeq its
negativity reads \beq
N_\phi(t)=\frac{1}{2}\sqrt{\left(\langle{S^2}\rangle(t)-1\right)^2}.
\eeq It is both formally invariant and numerical conserved only
under $H[2]\left(h_{03}=h_{30}\right)$, or only invariant in form
under $H[1,2]$. Here, we have defined that \beqa\label{h11}
H[1,1]&=&\frac{1}{2}\omega_1\left(\sigma_3\otimes\sigma_0+\sigma_0\otimes\sigma_3\right)
\nonumber\\
& &+g_1\left(\sigma_1\otimes\sigma_1+\sigma_2\otimes\sigma_2\right)+h_1\sigma_3\otimes\sigma_3,\\
\label{h12}
H[1,2]&=&\frac{1}{2}\omega_2\left(\sigma_3\otimes\sigma_0+\sigma_0\otimes\sigma_3\right)
\nonumber\\
&
&+f_1\left(\sigma_1\otimes\sigma_2-\sigma_2\otimes\sigma_1\right)+h_1\sigma_3\otimes\sigma_3.
\eeqa Fig.3 and Fig.4 show the surface graphics determined by
$\theta$, $\langle{S}^2\rangle(t)$ and $N_{\phi(\pm)}(t)$ under
$H[1,2]$.
\begin{figure}[h]%[tbp]
\begin{center}
\includegraphics[scale=0.56]{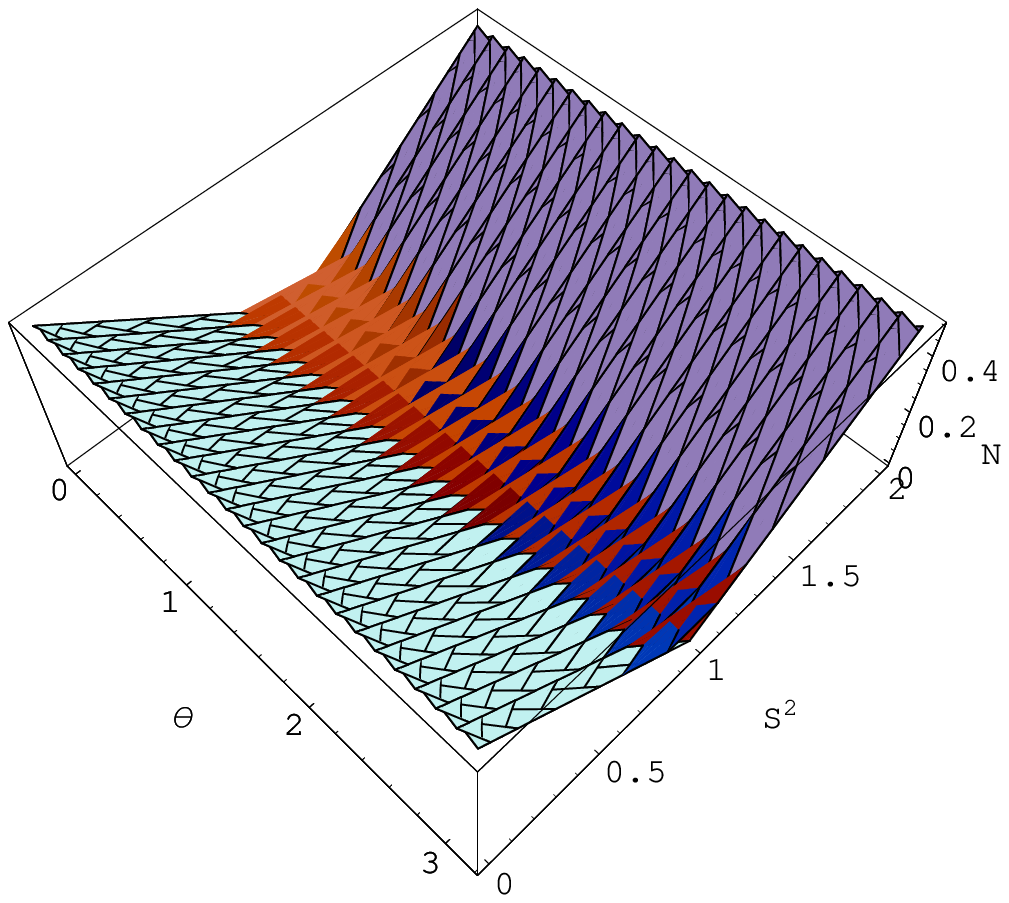}
\end{center} \vskip -0.2in
\caption{Under $H[1,2]$,
$\langle{S}^2\rangle_{\phi(+)}(t)=1+\sin^2\left(2T+2\theta_2\right)$,
$N_{\phi(+)}(t)=\sqrt{\sin^2\left(2T+2\theta_2\right)}/2$
($\alpha_2=0$, $T=2f_1$).}\label{mypic3}
\end{figure}
\begin{figure}[hbt]%[tbp]
\begin{center}
\includegraphics[scale=0.56]{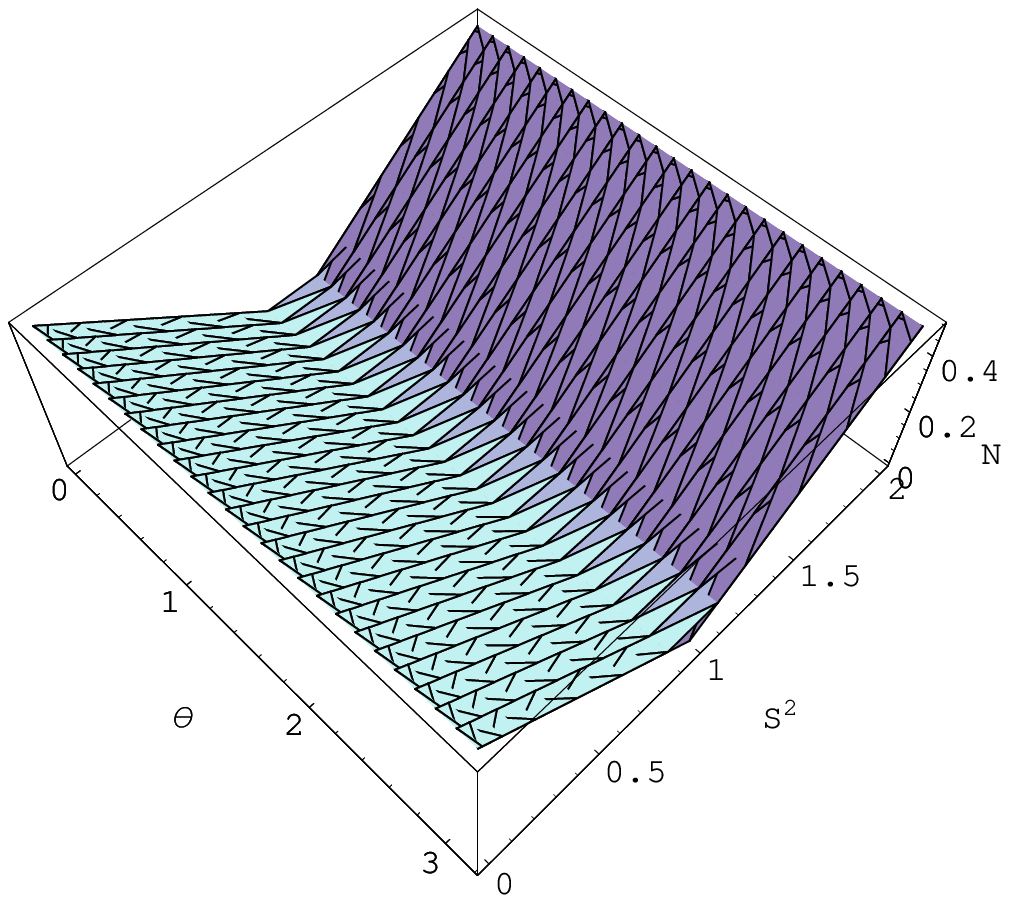}
\end{center} \vskip -0.2in
\caption{Under $H[1,2]$,
$\langle{S}^2\rangle_{\phi(-)}(t)=1+\sin^2\left(2T-2\theta_2\right)$,
$N_{\phi(-)}(t)=\sqrt{\sin^2\left(2T-2\theta_2\right)}/2$
($\alpha_2=\pi$, $T=2f_1$).}\label{mypic4}
\end{figure}

Similarly, we can obtain the invariance of the relations between the
entanglement and observables for the following six kinds of mixed
states at the initial time \beqa
\rho_1^M(0)&=&\sin^2\theta\ket{00}\bra{00}+\cos^2\theta\ket{11}\bra{11},\\
\rho_2^M(0)&=&\sin^2\theta\ket{01}\bra{01}+\cos^2\theta\ket{10}\bra{10},\\
\rho_3^M(0)&=&\sin^2\theta\ket{10}\bra{10}+\cos^2\theta\ket{11}\bra{11},\\
\rho_4^M(0)&=&\sin^2\theta\ket{01}\bra{01}+\cos^2\theta\ket{11}\bra{11},\\
\rho_5^M(0)&=&\sin^2\theta\ket{00}\bra{00}+\cos^2\theta\ket{10}\bra{10},\\
\rho_6^M(0)&=&\sin^2\theta\ket{01}\bra{00}+\cos^2\theta\ket{01}\bra{01}.
\eeqa They are separable, but they will become entangled after the
finite time evolution. Actually, these mixed states can be thought
of as the reduced density matrix from the compound systems initially
with entanglement. For example, the state
$\left(\sin\theta\ket{00}+\cos\theta\ket{11}\right)\otimes\ket{10}$
has entanglement between the first and second qubits. To partially
trace qubits 2 and 4, we will get $\rho_1^M(0)$ made of qubits 1 and
3, or partially trace qubits 1 and 3, we will get $\rho_3^M(0)$ made
of qubits 2 and 4, and so on. This implies that our methods and
following discussions can be similarly used to the cases of
entanglement transfer and with the environment-system interactions
\cite{Cavalcanti}.

Obviously, $N_{\rho^M_1}$ under both $H[2,1]$ and $H[2,2]$,
$N_{\rho^M_2}$ under $H[1,2]$ are, respectively, \beqa
N_{\rho_1^M}(t)&=&\frac{1}{2}\sqrt{\langle{S}_z\rangle(0)^2-\langle{S}_z\rangle(t)^2},\\
N_{\rho_2^M}(t)&=&\frac{1}{2}\sqrt{\left(\langle{S^2}\rangle(t)-1\right)^2}.\eeqa
While, under $H[1,2]$ \beqa
N_{\rho_i^M}(t)&=&\sqrt{\langle{S}_z\rangle_i^2(t)+\left(\langle{S}^2\rangle_i(t)
+\langle{S}_z\rangle_i(t)-
1\right)^2}\hskip 0.5cm \nonumber\\
& &+\langle{S}_z\rangle_i(t),\\
N_{\rho_j^M}(t)&=&\sqrt{\langle{S}_z\rangle_j^2(t)+\left(\langle{S}^2\rangle_j(t)
-\langle{S}_z\rangle_j(t)- 1\right)^2}\nonumber\\
& &-\langle{S}_z\rangle_j(t),\eeqa and under both $H[2,1]$ and
$H[2,2]$, \beqa
N_{\rho_k^M}(t)\!\!\!&=\!\!\!&\sqrt{\left(2\!-\!\langle{S}^2\rangle_k(t)\right)^2\!+\!
\left(\langle{S}^2\rangle_k(t)\!-\!
1\right)^2-\langle{S}_z\rangle_k^2(t)}\hskip 0.5cm\nonumber\\
& & +\langle{S}^2\rangle_i(t)-2, \eeqa where, $i=3,4$, $j=5,6$ and
$k=3,4,5,6$. The surface figures determined by $S_z$, $S^2$ and $N$
are Figs. 5 - 8
\begin{figure}[h]%[tbp]
\begin{center}
\includegraphics[scale=0.56]{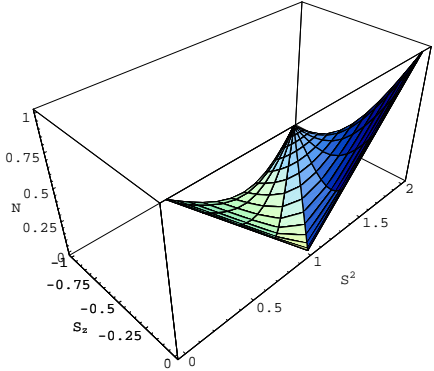}
\end{center} \vskip -0.2in
\caption{Under $H[1,2]$, $\langle{S}_z\rangle_i(t)=-\cos^2\theta_i$,
$\langle{S}^2\rangle_i(t)=\left[3+\cos(2\theta_i)
-(-1)^i\sin(2T)\sin^2\theta_i\right]/2$,
$N_{\rho_i^M}(t)=\sqrt{\cos^4\theta_i+\sin^2(2T)\sin^4\theta_i}-\cos^2\theta_i$
($i=3,4$ and $T=2f_2$).} \label{mypic5}
\end{figure}
\begin{figure}[h]%[tbp]
\begin{center}
\includegraphics[scale=0.56]{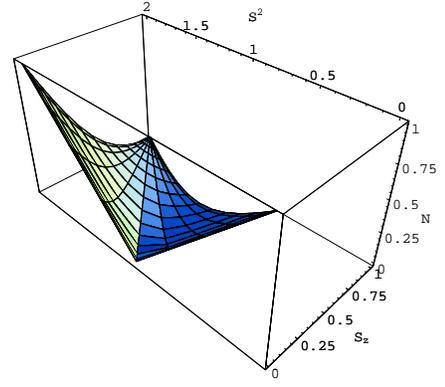}
\end{center} \vskip -0.1in
\caption{Under $H[1,2]$, $\langle{S}_z\rangle_j(t)=\sin^2\theta_j$,
$\langle{S}^2\rangle_j(t)=\left[3-\cos(2\theta_j)
-(-1)^i\sin(2T)\cos^2\theta_j\right]/2$,
$N_{\rho_j^M}(t)=\sqrt{\sin^4\theta_j+\sin^2(2T)\cos^4\theta_j}-\sin^2\theta_j$
($j=5,6$, $T=2 f_2 t$).}\label{mypic6}
\end{figure}
\begin{figure}[h]%[tbp]
\begin{center}
\includegraphics[scale=0.56]{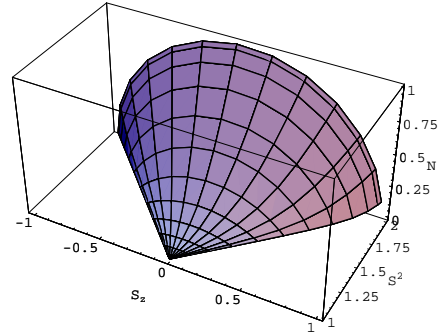}
\end{center} \vskip -0.2in
\caption{Under $H[2,1]$,
$\langle{S}_z\rangle_i(t)\!\!=\!\!-\cos(2T)\sin^2\theta_i$,
$\langle{S}^2\rangle_i(t)=\left[3+\cos(2\theta_i)\right]/2$,
$N_{\rho_i^M}(t)=\sqrt{\sin^4\theta_i+\sin^2(2T)\cos^4\theta_i}-\sin^2\theta_i$
($i=3,4$, $T=2 f_1 t$).} \label{mypic7}
\end{figure}
\begin{figure}[h]%[tbp]
%\vskip -0.1in
\begin{center}
\includegraphics[scale=0.56]{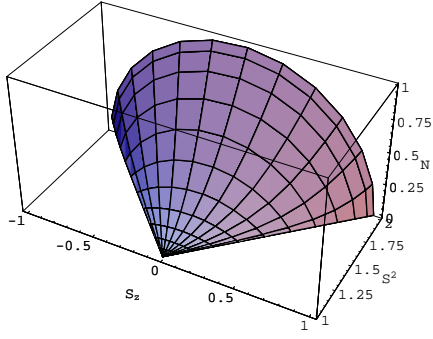}
\end{center} \vskip -0.2in
\caption{Under $H[2,2]$,
$\langle{S}_z\rangle_j(t)=\cos(2T)\sin^2\theta_j$,
$\langle{S}^2\rangle_j(t)=\left[3-\cos(2\theta_j)\right]/2$,
$N_{\rho_j^M}(t)=\sqrt{\cos^4\theta_j+\sin^2(2T)\sin^4\theta_j}-\cos^2\theta_j$
($j=5,6$, $T=2f_2 t$).} \label{mypic8}
\end{figure}

In summary, we propose a general scheme to seek for the relations
between entanglement and observables in principle, and we find and
verify such relations for the given state sets in two-qubits system,
that is, entanglement is expressed as the function of observables.
In addition, our methods can be similarly used to the cases of
entanglement transfer as well as with the environment-system
interaction. Because, Hamiltonian can be expressed by the involved
observables, our conclusions have implied the relation between
entanglement and energy. Moreover, our conclusions are more
determined relations than those derived out by only considering
energy. Then we obtain what Hamiltonian will keep their invariance
in form with time evolution. This is required to think that
entanglement is observable from our point of view. More important
and interesting thing in such a claim is that we demonstrate and
illustrate that the entanglement can be expressed as the functions
of a set of commutable observables $S^2$ and $S_z$ for eight kinds
of initial state sets in two-qubit systems. As well-known, the
involved observable should be measurable at the same time. Actually,
this means that entanglement can be measured by experiment in these
state sets. We expect it will be true in the near future.

We are grateful all the collaborators of our quantum theory group in
our university. This work was supported by the National Natural
Science Foundation of China under Grant No. 60573008.

%\vspace{-0.2in}

\end{document}